\documentclass{elsart}
\usepackage{graphics}

\newcommand{\aff}[2]{Dipartimento di Fisica dell'Universit\`a #1 e Sezione INFN, #2, Italy.}
\newcommand{\affd}[1]{Dipartimento di Fisica dell'Universit\`a e Sezione INFN, #1, Italy.}
\renewcommand{\to}        {\ensuremath{\rightarrow}}

\def\ifm#1{\relax\ifmmode#1\else$#1$\fi}
\def\pt#1,#2,{\ifm{#1\x10^{#2}}}  \def\x{\ifm{\times}}
\def\ab{\ifm{\sim}}   \def\f{\ifm{\phi}} \def\pio{\ifm{\pi^0\pi^0}}
\def\pic{\ifm{\pi^+\pi^-}}   \def\gam{\ifm{\gamma}}
\def\Rkspi{\ifm{R_S^\pi}}   \def\epm{\ifm{e^+e^-}}

\newcommand{\Dafne}       {DA\char8NE}
\newcommand{\kzero}       {\ensuremath{K^{0}}}
\newcommand{\kzerob}      {\ensuremath{\bar{K}^{0}}}
\newcommand{\ks}          {\ensuremath{K_{S}}}
\newcommand{\kl}          {\ensuremath{K_{L}}}
\newcommand{\tzero}       {\ensuremath{T_{0}}}
\newcommand{\kspopo}      {\ensuremath{\ks\!\to\!\pi^{0}\pi^{0}}}
\newcommand{\kspppm}      {\ensuremath{\ks\!\to\!\pi^{+}\pi^{-}}}
\newcommand{\klpc}        {\ensuremath{\kl\!\to\!\pi^{+}\pi^{-}}}
\newcommand{\klmm}        {\ensuremath{\kl\!\to\!\mu^{+}\mu^{-}}}

\newcommand{\bstar}       {\ensuremath{\beta^{\ast}}}
\newcommand{\gampiumeno}  {\ensuremath{\Gamma(\kspppm)}}
\newcommand{\gampiumenog} {\ensuremath{\Gamma(\kspppm(\gamma))}}
\newcommand{\gamzerozero} {\ensuremath{\Gamma(\kspopo)}}
\newcommand{\phipppmpo}   {\ensuremath{\phi\!\to\!\pi^{+}\pi^{-}\pi^{0}}} 
\newcommand{\phikskl}     {\ensuremath{\phi\!\to\!\ks\kl}}
\newcommand{\eV}{{e\kern-.07em V}}
\newcommand{\MeV}{{\rm \,M\eV}}
\newcommand{\GeV}{{\rm \,G\eV}}
\newcommand{\pb}{{\rm \,pb}}
\newcommand{\ps}{{\rm \,ps}}
\newcommand{\ns}{{\rm \,ns}}
\newcommand{\mm}{{\rm \,mm}}
\newcommand{\m}{{\rm \,m}}
\newcommand{\cm}{{\rm \,cm}}
\newcommand{\um}{\,\textrm{$\mu$m}}
\newcommand{\T}{{\rm \,T}}

\newcommand{\Fig}{Fig.}
\newcommand{\Ref}{Ref.}
\newcommand{\Tab}{Tab.}
\begin{document}

\begin{frontmatter}
\title{Measurement of $\gampiumenog/\gamzerozero$}
\collab{The KLOE Collaboration}
%\author[Roma2]{M.~Adinolfi},
\author[Na] {A.~Aloisio},
\author[Na]{F.~Ambrosino},
%\author[Frascati,Moscow]{A.~Andryakov},
\author[Frascati]{A.~Antonelli},
\author[Frascati]{M.~Antonelli},
%\author[Frascati]{F.~Anulli},
\author[Roma3]{C.~Bacci},
\author[Frascati]{R.~Baldini-Ferroli},
%\author[Trieste]{G.~Barbiellini},
%\author[Roma3]{F.~Bellini},
\author[Frascati]{G.~Bencivenni},
\author[Frascati]{S.~Bertolucci},
\author[Roma1]{C.~Bini},
\author[Frascati]{C.~Bloise},
% ...
%\author[Beijing]{Y.~Wu},
%\author[Beijing]{Y.~G.~Xie},
%\author[Pisa]{P.~F.~Zema}
%\author[Beijing]{P.~P.~Zhao},
%\author[Frascati]{Y.~Zhou}
%%%%
\address[Bari]{\affd{Bari}}
\address[Virginia]{Physics Department, University of Virginia, Charlottesville, VA, USA.}
\address[Frascati]{Laboratori Nazionali di Frascati dell'INFN, Frascati, Italy.}
\address[Karlsruhe]{Institut f\"ur Experimentelle Kernphysik, Universit\"at Karlsruhe, Germany.}
\address[Lecce]{\affd{Lecce}}
\address[Na]{Dipartimento di Scienze Fisiche dell'Universit\`a ``Federico II'' e Sezione INFN, Napoli, Italy}
\address[Energ]{Dipartimento di Energetica dell'Universit\`a ``La Sapienza'', Roma, Italy.}
\address[Roma1]{\aff{``La Sapienza''}{Roma}}
\address[Roma2]{\aff{``Tor Vergata''}{Roma}}
\address[Roma3]{\aff{``Roma Tre''}{Roma}}
\address[Pisa]{\affd{Pisa}}
\address[StonyBrook]{Physics Department, State University of New York at Stony Brook, NY, USA.}
\address[Trieste]{\affd{Trieste}}
\address[Beijing]{Permanent address: Institute of High Energy Physics, CAS, Beijing, China.}
\address[Moscow]{Permanent address: Institute for Theoretical and Experimental Physics, Moscow, Russia.}
%\address[Negev]{Physics Department, Ben-Gurion University of the Negev, Israel.}
%\address[Columbia]{Physics Department, Columbia University, New York, USA.}
\corauth[cor1]{Corresponding author: Matteo Palutan
INFN - LNF, Casella postale 13, 00044 Frascati (Roma), 
Italy; tel. +39-06-94032697, e-mail matteo.palutan@lnf.infn.it}
\corauth[cor2]{Corresponding author: Tommaso Spadaro
INFN - LNF, Casella postale 13, 00044 Frascati (Roma), 
Italy; tel. +39-06-94032696, e-mail tommaso.spadaro@lnf.infn.it}
\corauth[cor3]{Corresponding author: Paolo Valente
INFN - LNF, Casella postale 13, 00044 Frascati (Roma), 
Italy; tel. +39-06-94032761, e-mail paolo.valente@lnf.infn.it}
\begin{abstract}
We have measured the ratio $\Rkspi\!=\!\gampiumenog/\gamzerozero$ with the KLOE 
detector at the \Dafne\ $e^+e^-$ collider. This measurement is fully inclusive with
respect to the $\pi^{+}\pi^{-}\gamma$ final state.
%Kaons from \ab\pt5,7, \f\ mesons produced at the \Dafne\ $e^+e^-$. 
The sample of over $10^6$ two-pion decays of tagged \ks\ mesons allows 
a statistical error as low as $\ab\!0.1\%$ to be obtained. 
The accuracy is limited by systematic uncertainties, which are
estimated primarily from data. 
We find $\Rkspi\!=\!2.236\!\pm\!0.003_{\rm stat}\!\pm\!0.015_{\rm syst}.$
\end{abstract}
\end{frontmatter}
%
%\section{Introduction}
%
The most recent measurement of $\Rkspi\!=\!\gampiumeno/\gamzerozero$ dates back to 1976. 
The authors quote an accuracy of 4.3\%~\cite{Everhart:1976ub}. 
This and other results of similar accuracy determine the current world-average value
of \Rkspi, $2.197\!\pm\!0.026,$ the fractional error on which is 1.2\%.
%Groom \textit{et al.}~\cite{pdg2000} quote an average for \Rkspi\ of  \idemest, a
%fractional error of  
Moreover, the averaging of the existing results is 
somewhat questionable, since the various experiments have not clearly described their 
procedures for handling radiative \pic\gam\ decays.
%{\it slightly} objectionable since the various experiments have removed radiative 
%\pic\gam\ decays by approximate and 
%varying cuts of unknown efficiency.
It is currently of interest to improve the accuracy on \Rkspi, with particular
attention to an accurate 
determination of what part of the radiative spectrum is included in the quoted 
result~\cite{Cirigliano}.
%Progress in chiral perturbation~\cite{gasser,bijnens} requires the improved accuracy. 
The validity of the $\Delta I\!=\!1/2$ rule, empirically established in 
$K$, $\Lambda$, $\Sigma$, and $\Xi$ decays, has never been satisfactorily understood. 
This lack of understanding is closely connected with the difficulty in 
relating $\Re(\epsilon'/\epsilon)$ to the CKM matrix elements~\cite{CiuchiniM}.
The value of \Rkspi\ also enters into the theoretical calculation for the ratio
of branching ratios for the decays \klmm\ and \klpc~\cite{Littenberg:2002sx}.

We describe the first measurement of \Rkspi\ to achieve an accuracy 
below 1\%, which moreover fully includes \pic\gam\ radiative decays. 
This should allow the extraction of the isospin $I\!=\!0$ and $I\!=\!2$ 
decay amplitudes. 
%Since in none of the previous measurements of $\Gamma(K\to\pi^{+}\pi^{-})$ the treatment of the radiation of photons 
%is specified, extraction of the amplitudes from them is affected by a larger uncertainty than that due to experimental 
%errors~\cite{Cirigliano}.%

The data were collected with the KLOE~\cite{kloep,kloet} detector at
\Dafne, the Frascati \f-factory~\cite{Dafneref}. \Dafne\ is an \epm\ collider which
operates at a center of mass energy 
of \ab1020\MeV, the mass of the \f-meson. 
In the following, we use a coordinate system with 
the $z$ axis along the nominal beam line, the $x$ axis in the horizontal plane
toward the machine center, the $y$ axis vertical, and the origin 
at the center of the collision region. 
Equal-energy positron and electron beams collide at an angle of 25\,mrad, producing \f-mesons 
nearly at rest: $p(\phi)_{y,z}\!=\!0$ and $p(\phi)_x\!=\!12.5 \MeV/c.$ 
\f-mesons decay \ab34\% 
of the time into nearly collinear \kzero\kzerob\ pairs. Because $J^{PC}(\f)\!=\!1^{--},$ 
these pairs are in an antisymmetric state, so that the final state is always 
\ks\kl. Two photon intermediate states can lead
 to \ks\ks\ and \kl\kl\ final states, but this contamination is 
%less than $\ab10^{-9}$
negligible~\cite{Dunietz:1987jf,brownclose}. 
All of the above implies that the detection of a \kl\ 
guarantees the presence of a \ks\ of given momentum and direction.
This fact can be exploited to identify \ks-mesons independent of their decay mode.
We refer to this technique as \textit{\ks\ tagging}.
%
%\section{Experimental setup}
%

The KLOE detector consists of a large cylindrical drift chamber, DC, surrounded by a 
lead-scintillating fiber sampling calorimeter, EMC. A superconducting coil
surrounding the calorimeter provides a 0.52\T\ magnetic field. 

The drift chamber~\cite{DCnim}, 4\m\ in diameter and 3.3\m\ long, 
has 12\,582 all-stereo tungsten 
sense wires and 37\,746 aluminum field wires.
The chamber shell is made of carbon fiber-epoxy composite and the gas used is a 
90\% helium, 10\% isobutane mixture. 
These features maximize transparency to photons and reduce $\kl\!\to\!\ks$ regeneration and 
multiple scattering. 
The position resolutions are $\sigma_{xy}\approx 150\um$ 
and $\sigma_z\approx 2\mm.$ The momentum resolution is 
$\sigma(p_{\perp})/p_{\perp}\approx 0.4\%.$ Vertices are 
reconstructed with a spatial resolution of $\approx 3\mm.$

The calorimeter~\cite{EmCnim} is divided into a barrel and two endcaps containing a 
total of 88 modules, and covers 98\% of the solid angle. 
The modules are read out at both ends by photomultipliers. 
The arrival times of particles and the three-dimensional positions of 
the energy deposits are determined from the signals at the two ends. 
The readout granularity is 
$\ab4.4\!\times\!4.4\cm^2$; the 2440 ``cells'' are arranged in layers five-deep.
Cells close in time and space are grouped into a ``calorimeter cluster.''
For each cluster, the energy $E_{\rm CL}$ is the sum of the cell energies, and
the time $t_{\mathrm{CL}}$ and position $\vec{r}_{\mathrm{CL}}$ are calculated as
energy-weighted averages over the fired cells. The energy and time resolutions 
are $\sigma_E/E \!=\! 5.7\%/\!\sqrt{E (\GeV)}$ and  
$\sigma_t\!=\!54\ps/\!\sqrt{E (\GeV)}\oplus50\ps,$ respectively.

While the trigger~\cite{TRGprop} uses calorimeter and chamber information, only the 
calorimeter 
trigger, which requires two local energy deposits above threshold 
($50$\MeV\ on the barrel and $150$\MeV\ on the endcaps),
is used for the present measurement.
Recognition and rejection of cosmic-ray events are also performed at the trigger level. 
Events with two energy deposits above a 30\MeV\ threshold in the outermost 
calorimeter plane are identified as cosmic-ray events and rejected.
The trigger has a large time spread with respect to the beam crossing time. However, it is 
synchronized with the machine RF divided by 4, $T_{\rm sync}\!=\!10.8\ns,$ 
with an accuracy of 50\ps. 
The time \tzero\ of the bunch crossing producing an event is determined after 
event reconstruction. 
%
%\section{\ks\ tagging}
%

The mean decay lengths of the \ks\ and \kl\ are $\lambda_{S}\sim0.6\cm$ and
$\lambda_{L}\sim350\cm.$ About 50\% of \kl's 
therefore reach the calorimeter before decaying. 
The \kl\ interaction in the calorimeter (\textit{\kl-crash}) is identified 
by requiring a cluster of energy above 
200\MeV, which is not associated with any track, and whose time corresponds to a velocity 
$\beta\!=\!r_{\mathrm{CL}}/ct_{\mathrm{CL}}$
compatible with that of the \kl, $\beta(\kl)\!\ab\!0.214.$ These \kl-crash
events are used for \ks\ tagging. The tagging efficiency depends slightly
on the \ks\ decay mode:
the ratio $R_{\mathrm{tag}}$ of the tagging efficiencies for \kspppm\ and \kspopo\
is determined from data.

\begin{figure}[ht]
  \begin{center}
    \mbox{
     \resizebox{!}{.6\textwidth}{\includegraphics{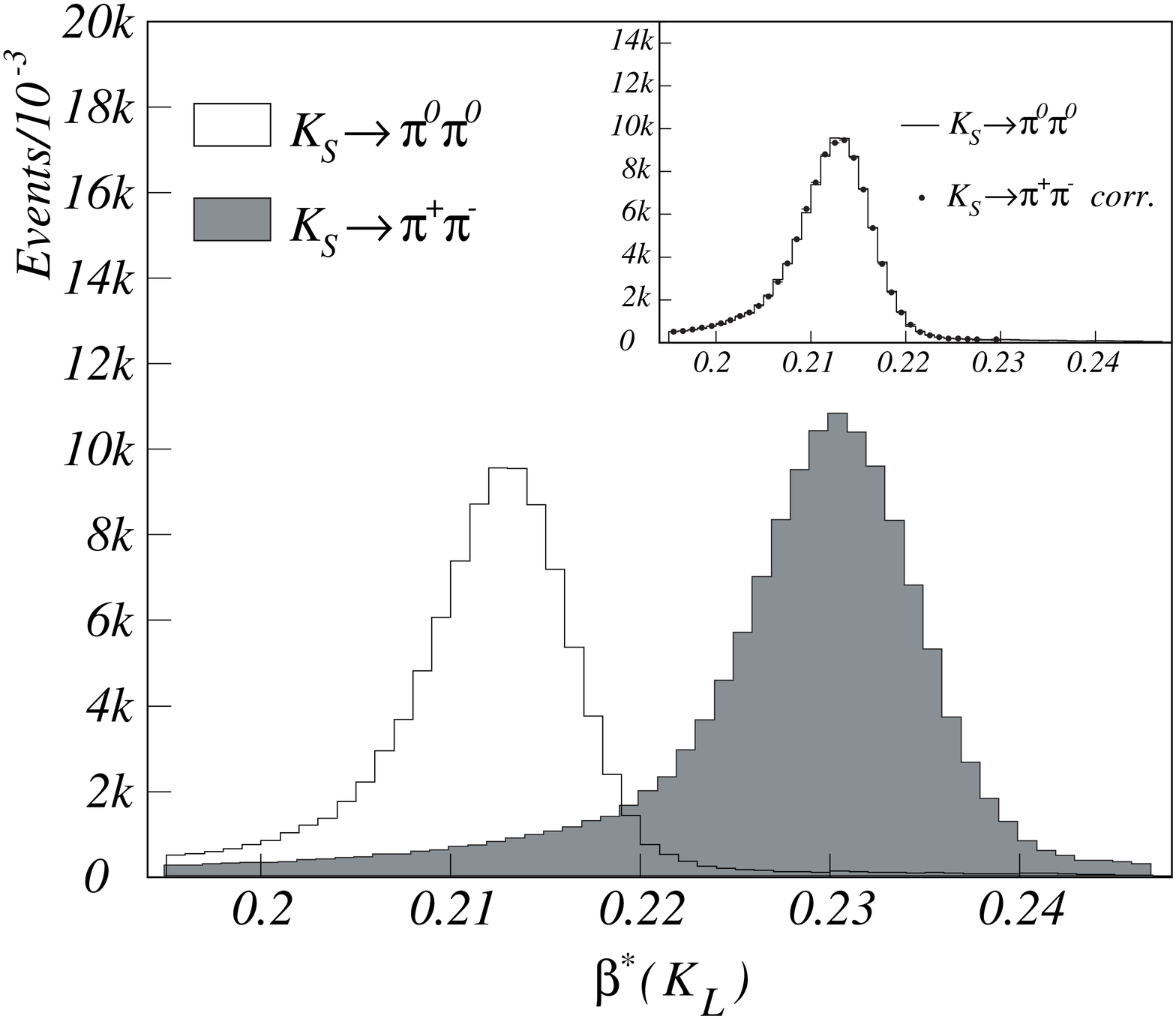}}
     }
    \caption{Distribution of the \kl\ velocity in the $\phi$-rest frame,
     for \kspopo\ (empty histogram) and \kspppm\ (shaded histogram) events. 
     In the inset, the distribution for
     \tzero-corrected \kspppm\ events (dots) is superimposed on that 
     for \kspopo\ events (line).}
    \label{betastar}
  \end{center}
\end{figure}

%\red a charged track, \blue and with a ``cluster velocity'' ${\bf v}_{\rm CL}={\rm d}{\bf 
%r}/{\rm d}t$ which -- since in fact the cluster does not move at all -- we redefine as 
%$|\vec r_{\mathrm{CL}}|/ct_{\mathrm{CL}}$ \red it should really be $|\vec r_{\mathrm{CL}
%}|/(ct_{\mathrm{CL}})$, \green
%compatible with that of the \kl, completely ignoring the fact that $\vec r_{\rm CL}$ 
%was never defined. \blue -- it is now. \red The point is that 1. $ {\rm d}{\bf r}/{\rm d}t$ 
%is independent of the choice of frame, while $|\vec r|/t$ is not and 2. the cluster does 
%not move, even if you use ``.....''!!! \black
The interaction time \tzero, which must be known for the measurement of the cluster
times, is obtained from the first particle reaching the calorimeter (pions or photons for 
the events of interest) assuming a velocity $\beta\!=\!1.$
In order to reduce the probability that \tzero\ is accidentally determined from 
a particle arising from machine background, 
we require the \tzero\ to be fixed by a cluster with 
energy $E_{\rm CL}\!>\!50$\MeV\ and distance to the beam line $\rho_{\rm CL}\!>\!60\cm.$
Charged pions from \kspppm\ arrive at the EMC $\ab\!3\ns$ 
later than \gam's from $\pi^0$ decays. The computed time \tzero\ is therefore 
delayed by one RF period, $\ab\!2.7\ns.$ This results
in a mismeasurement of the velocity for long-lived kaons accompanied by 
\kspppm\ decays. 
The distribution of the observed \kl\ velocity transformed to the $\phi$-rest frame,
$\bstar,$ is shown in \Fig~\ref{betastar}. 
The effect of the incorrect \tzero\ assignment for the case of \kspppm\ is evident.
%In order to let the tagging technique be independent of 
%the actual \ks\ decay mode, 
%the above effect is not corrected at this level of the analysis;
Events are selected by requiring $0.195\!\leq\!\bstar\!\leq\!0.2475.$ 
This window in \bstar\ encompasses both peaks and gives a
tagging efficiency of $\ab\!20\%,$ which differs slightly for each of the two \ks\ 
decay modes. The ratio of tagging efficiencies $R_{\mathrm{tag}}$ is evaluated from a 
subsample of \kspppm\ events in which both pion tracks are associated to a calorimeter 
cluster. In this case, the event \tzero\ can be correctly evaluated making use of the 
measured pion track lengths and velocities, resulting in the \bstar\ spectrum shown in
the inset of \Fig~\ref{betastar}, which is very similar 
to that for \kspopo\ events. This corrected \bstar\ distribution and that obtained 
from the standard \tzero\ estimate are normalized to the same number of \kl\ interactions. 
$R_{\mathrm{tag}}$ is computed from the ratio of the total number of events inside 
the \bstar\ window in each of the two cases. The effects of residual differences 
between the \bstar\ distributions 
for \kspopo\ and \tzero-corrected 
\kspppm\ events are included in the determination of the systematic error.

Other sources of bias in the evaluation of $R_{\mathrm{tag}}$ have been taken into account
as well. 
The probability that the event contains a cluster which determines \tzero\ is slightly 
different for \pic\ and \pio\ decays. 
These probabilities are estimated using \ks\ events tagged by  
 $\kl\to\pi^{+}\pi^{-}\pi^{0}$ decays. For the identification of such events, 
knowledge of \tzero\ is not
necessary. These events are identified by searching for two tracks forming a 
vertex in the drift chamber volume. 
The missing momentum of the $\pi^{0}$ is determined by kinematic closure at the \kl\ vertex.
Photons from the $\pi^{0}$ are 
identified using \textit{the difference} between their times of flight 
and by requiring that the 
$\gamma\gamma$ pair be back-to-back in the $\pi^{0}$ rest frame. 

Accidental overlap of a $\pi^{\pm}$ track with 
the \kl\ cluster produces a loss of \kspppm\ events. The magnitude of this effect has been 
estimated using \kl-crash events selected without requiring that the cluster not be associated
with any track. 

Finally, a small fraction of \phikskl\ events is rejected by the cosmic-ray veto, 
since some \kl-crash interactions can deposit energy in the outermost EMC plane.
$R_{\mathrm{tag}}$ is therefore corrected by the ratio of the inefficiencies induced
by the cosmic-ray veto for \kspppm\ 
and \kspopo\ events. This ratio is obtained using a subsample of 
events for which the cosmic-ray veto was recorded but not 
enforced.

All of the above effects result in corrections to $R_{\mathrm{tag}}$, which are listed 
in \Tab~\ref{effitab}. The largest uncertainty is due to the correction for the
\bstar\ cut.

\begin{figure}[ht]
  \begin{center}
    \mbox{
     \resizebox{!}{5cm}{\includegraphics{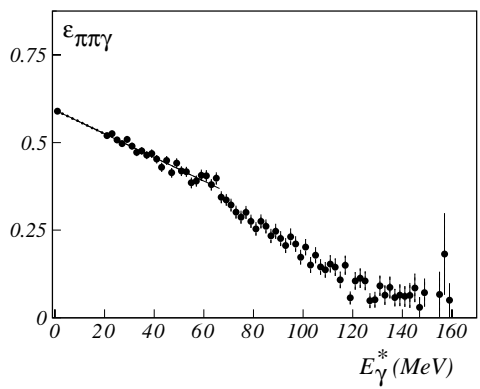}}
     }
    \caption{Efficiency for \ks\to\pic\gam\ events vs radiated energy
    in the \ks\ rest frame $E^{\ast}_{\gamma},$ from Monte Carlo simulation.}
    \label{pipigi}
  \end{center}
\end{figure}

\ks\to\pic(\gam) events are selected by requiring the presence of
two tracks of opposite charge that intersect a cylinder 4\cm\ in radius and 10\cm\ in length
along the $z$-axis, centered at the interaction region (IR). 
Pion momenta and polar angles must satisfy the fiducial cuts 
$120\!\leq\!p\!\leq\!300\MeV/c$ and  
$30^\circ\!\leq\!\theta\!\leq\!150^\circ.$ 
Pions must also reach the EMC, as opposed to spiralling in the drift chamber.
The requirement that both tracks reach the EMC results in an acceptance which depends on 
the energy of the photon in radiative decays. 
\Fig~\ref{pipigi} shows the dependence of 
the efficiency $\epsilon_{\mathrm{\pi\pi\gamma}}$ on the $\gamma$ energy in the \ks\ rest
frame $E^\ast_\gamma$, obtained by Monte Carlo 
(MC) simulation for \ks\to\pic\gam\ events with 
$E^\ast_\gamma>20$\MeV\ and \kspppm\ events 
without radiation ($E^\ast_\gamma\!=\!0$).
The efficiency for $0\!<\!E^\ast_\gamma\!<\!20\MeV$ is obtained
by linear interpolation. The overall acceptance is obtained by folding the above efficiency 
with the photon spectrum from \Ref~\cite{Cirigliano}.
The result differs from that computed assuming no radiation by $\sim0.3\%.$

The single-track reconstruction efficiency is obtained in bins of 
$(p,\theta)$ using \kspppm\ events identified by a single reconstructed pion track. 
This selection is applied to both data and MC. 
The ratio of data and MC reconstruction efficiencies is found to be constant 
($0.990\pm0.001$) for pions with momenta and polar angles the satisfy the fiducial
cuts used in the analysis. 
The overall MC acceptance is therefore scaled accordingly and used to correct the data. 
Systematic errors 
due to possible misalignments of the DC are evaluated by studying the dependence of 
the acceptance on the angular cut applied and are found to be negligible.
The acceptance is listed in \Tab~\ref{effitab}.

The trigger efficiency is also obtained from data. \kl\ interactions always fire at least 
one trigger sector. About 40$\%$ of the time, however, the \kl-crash fires two nearby 
sectors, thus producing a valid trigger by itself. 
We use these events to find the probability that 
at least one \ks\ decay product complements the \kl-crash cluster to satisfy the trigger
condition in the remaining 60\% of the events. The trigger efficiency is given in 
\Tab~\ref{effitab}.
 
%
%\section{Selection of \kspopo\ events}
%
\kspopo\ events are identified by the prompt photon clusters from $\pi^{0}$ decays. 
A prompt photon cluster must satisfy 
$|t_{\mathrm{CL}}-r_{\mathrm{CL}}/c|\!\le\!5\sigma_t,$
where $\sigma_t$ is the energy-dependent time resolution, and must not be associable 
to any track. Machine background is reduced by cuts on energy and polar angle; we require
$E_{\rm CL}\!>\!20\MeV$ and $\cos\theta\!<\!0.9.$
Three or more prompt photons are required to accept a \kspopo\ event.

The photon-detection efficiency in bins of energy and polar 
angle has been obtained from data using a sample of \phipppmpo\ events. 
The acceptance for \kspopo\ events is obtained from MC simulation 
and is listed in \Tab~\ref{effitab}.
The simulation incorporates the experimentally determined photon-detection efficiencies;
we rely on it essentially for the averaging over event geometries and kinematics.
The 0.2$\%$ error on the \kspopo\ selection efficiency is dominated by the uncertainty 
in the photon-detection efficiency below 50\MeV\ 
due to non-linearity in the energy response of the 
calorimeter and/or the photon reconstruction algorithm.  
The trigger efficiency, also listed in \Tab~\ref{effitab}, is evaluated by the same 
method used for \kspppm\ events. 

Background in the sample of \kspopo\ events is mainly due to \kspppm\ events in 
which the track-to-cluster 
association failed, and in which there were residual split and accidental 
clusters. This contamination induces an error on the counting of events with three or more
prompt clusters that is lower 
than 0.1\%, as estimated directly from data. 
\begin{table}[ht]
  \begin{center}
  \begin{tabular}[c]{c|c}\hline
 Source & $R_{\mathrm{tag}}$ \\ \hline\hline
     $\bstar$ window               & $0.986\pm 0.005$ \\ \hline
          \tzero\ efficiency       & $0.990\pm 0.001$ \\ \hline
         Track-cluster association & $0.998\pm0.002$  \\ \hline
         Cosmic-ray veto           & $1.003\pm0.002$  \\ \hline \hline
         Total                     & $0.977\pm0.005$  \\ \hline \hline
\noalign{\vglue2mm}\hline
                                           & $\epsilon(\pi^+\pi^-)$  \\ \hline\hline
 Event selection                       
& $0.5760\pm 0.0015$  \\ \hline
 Trigger                                   & $0.989\pm 0.001$    \\ \hline
 Total                                     & $0.5697\pm 0.0016$  \\ \hline\hline
\noalign{\vglue2mm}\hline 
                                           & $\epsilon(\pi^0\pi^0)$   \\ \hline\hline
 Event selection                           & $0.9005\pm 0.0018$  \\ \hline
 Trigger                                   & $0.9986\pm 0.0004$  \\ \hline
 Total                                     & $0.8992\pm 0.0018$  \\ \hline\hline
  \end{tabular}
  \end{center}
  \caption{The analysis steps are listed, together with the associated efficiencies.}
  \label{effitab}
\end{table}
%
%\section{Result}
%

The present measurement of \Rkspi\ is based on an integrated luminosity of 
$17\pb^{-1}$ collected during the year 2000. The data set contains
$N^{\pm}=1\,060\,821$ selected \kspppm\ events
and $N^{\mathrm{00}}=766\,308$ selected \kspopo\ events. 
The value for \Rkspi\ is obtained by correcting the ratio of these numbers for the ratio 
of tagging and selection (including trigger) efficiencies:
$$
\Rkspi\ = \frac{N^{\pm}}{N^{\mathrm{00}}}  \times \frac{1}{R_{\mathrm{tag}}} \times
\frac{\epsilon_{\mathrm{sele}}^{\mathrm{00}}}{\epsilon_{\mathrm{sele}}^{\pm}} 
$$
The result is 
$$
2.236\pm0.003_{\mathrm{stat}}\pm0.015_{\mathrm{syst}}
$$ 
The contributions to the estimated errors are listed in \Tab~\ref{systab}. 
Additional details are available in \Ref~\cite{pipinote}. 
Our result is consistent with but slightly higher than  
the current world-average value $\Rkspi=2.197\pm0.026$~\cite{pdg2000}.
This might be expected, since our measurement was designed to
include the entire radiative spectrum.
\begin{table}[ht]
  \begin{center}
  \begin{tabular}[c]{c|c}\hline
        Source                             & Error, \% \\ \hline\hline
        Event count                        &  0.14   \\ \hline 
        Tracking                           &  0.26   \\ \hline
        Cluster counting                   &  0.20   \\ \hline
        Trigger                            &  0.10   \\ \hline
        Ratio of tag efficiencies          &  0.55   \\ \hline
        Cosmic-ray veto                    &  0.21   \\ \hline\hline
       Total error                         &  0.70   \\ \hline
  \end{tabular}                          
  \end{center}
  \caption{Contributions to the fractional error on \Rkspi.}
  \label{systab}
\end{table}

%
%\section{Acknowledgments}
%
We thank the \Dafne\ team for their efforts in maintaining low
background running conditions and their collaboration during all
data-taking. We also thank F. Fortugno for his efforts in ensuring
good operations of the KLOE computing facilities. This work was
supported in part by DOE grant DE-FG-02-97ER41027; by 
EURODAPHNE contract FMRX-CT98-0169; by the German Federal Ministry of Education 
and Research (BMBF) contract 06-KA-957; 
by Graduiertenkolleg 'H.E. Phys. and Part. Astrophys.' of 
Deutsche Forschungsgemeinschaft contract No. GK 742;
by INTAS contracts 96-624 and 99-37; and by TARI contract HPRI-CT-1999-00088.
\bibliographystyle{elsart-num}
% latex file of the bibliography (draft_ksppoo.bbl, obtained from bibtex draft_ksppoo)
\bibliography{Ksppoo}

\end{document}